# Quantum Feasibility Labeling for NP-complete Vertex Coloring Problem

**Junpeng Zhan[1], Member, IEEE**

[1]Department of Renewable Energy Engineering, Alfred University, Alfred, NY 14802 USA

Corresponding author: Junpeng Zhan (e-mail: zhanj@alfred.edu).

This research was supported in part by the NSF ERI program, under award number 2138702.

**ABSTRACT** Many important science and engineering problems can be converted into NP-complete problems which are of significant importance in computer science and mathematics. Currently, neither existing classical nor quantum algorithms can solve these problems in polynomial time. To address this difficulty, this paper proposes a quantum feasibility labeling (QFL) algorithm to label all possible solutions to the vertex coloring problem, which is a well-known NP-complete problem. The QFL algorithm converts the vertex coloring problem into the problem of searching an unstructured database where good and bad elements are labeled. The recently proposed variational quantum search (VQS) algorithm was demonstrated to achieve an exponential speedup, in circuit depth, up to 26 qubits in finding good element(s) from an unstructured database. Using the labels and the associated possible solutions as input, the VQS can find all feasible solutions to the vertex coloring problem. The number of qubits and the circuit depth required by the QFL each is a polynomial function of the number of vertices, the number of edges, and the number of colors of a vertex coloring problem. We have implemented the QFL on an IBM Qiskit simulator to solve a 4-colorable 4-vertex 3-edge coloring problem.

## I. INTRODUCTION

NP-complete problems [1], [2] are a type of problem for which it is believed that there is no efficient algorithm for solving them. Many science and engineering problems can be converted into NP-complete problems. These problems are typically characterized by the need to find the optimal solution among a large number of possible solutions.

Here are a few examples of the many science and engineering problems that can be converted into NP-complete problems: Traveling Salesman Problem [3]–[5] (of interest in transportation and logistics), Knapsack Problem [6]–[9] (of interest in resource allocation and inventory management), Satisfiability Problem [10]–[14] (SAT, of interest in computer science, artificial intelligence, and logic), Subset Sum Problem [1] (of interest in cryptography and computer security), Graph Coloring Problem [15], [16] (of interest in scheduling and resource allocation), Hamiltonian Cycle Problem [15]–[17] (of interest in network design), and the Steiner Tree Problem [17] (of interest in telecommunications and computer networks).

Despite the many efforts [3]–[17] to develop classical algorithms for solving NP-complete problems, it is generally believed that there is no classical algorithm that can solve NP-complete problems in polynomial time, which means that the time and resources required to solve these problems increase faster than a polynomial function of the size of the problem. Whether NP-complete problems can be solved efficiently is an extremely important but unsolved question. The Clay Mathematics Institute has offered a prize of $1 million for a solution to the problem.

Quantum computing has been the subject of intense research and development in recent years due to its potential to solve certain types of problems much faster than classical computers [18]–[29]. The progress in quantum computer hardware development has been particularly impressive in recent years. Researchers have made significant advances in the design and construction of quantum computer hardware [27], [28], [30]–[33], and several companies have released commercial quantum computers that are available for use by researchers and businesses [34]–[36]. These quantum computers have demonstrated impressive performance on a variety of tasks [18], [27], [28], and their capabilities are expected to continue to improve in the coming years [37], [38].

In addition to the progress in quantum computer hardware development, there has also been significant progress in the development of a wide range of quantum algorithms for a variety of different types of problems [20], [21], [29], [39]–[49]. Some of these algorithms have demonstrated impressive performance on tasks that are difficult or impossible to solve using classical algorithms [18], [27], and it is expected that the development of quantum algorithms will continue to be an active area of research in the coming years.

Depending on whether classical computers are used, we can divide quantum algorithms into 1) pure quantum algorithms such as Shor's algorithm for factorization [24] and Grover's algorithm for searching unordered lists [22], [23], and 2) variational quantum algorithms (VQAs) [50]–[52] which involve both classical and quantum computers. VQAs have been successfully used in areas such as optimization problems [53]–[55], quantum chemistry [32], [50], [52], [56], [57], and machine learning [42], [58]–[63], and are expected to help achieve systematic quantum supremacy over classical algorithms using hundreds of qubits which are already available [34].

Given the promising potential and recent success of quantum computing, it is natural to ask whether it can provide a quantum exponential speedup in solving NP-complete problems. *Unfortunately, it is widely believed (though not proven) that the answer is no* [64], [65]. However, a recent paper [29] proposed a variational quantum search (VQS) algorithm, which has been shown to have an exponential advantage, in the circuit depth, over Grover's search algorithm in searching an unstructured database, up to a limit of 26 qubits (the limit is due to the memory constraints of the 48-GB GPU used in the calculation). According to paper [29], a depth-10 quantum circuit can amplify the total probability of $k$ ($k \geq 1$) good elements, out of $2^n$ elements represented by $n$+1 qubits, from $k/2^n$ to nearly 1 for $n$ up to 26. Additionally, the maximum depth of quantum circuits in the VQS increases linearly with the number of qubits. Given that Grover's algorithm can provide quadratic speedup in solving NP-complete problems, the VQS could solve these problems in polynomial time for up to 26 qubits. However, for larger instances beyond 26 qubits, the efficiency of VQS needs further in-depth exploration and investigation.

One of the fundamental properties of NP-complete problems is that they can be reduced to one another through polynomial-time transformations. This means that if an efficient (polynomial-time) algorithm exists for solving one NP-complete problem, it can be adapted to solve all NP-complete problems. A breakthrough in one problem (such as vertex coloring problem) can lead to advances in solving others (such as SAT or Hamiltonian Cycle).

This paper focuses on the vertex coloring problem [15], [16], which is a well-known NP-complete problem in graph theory. It involves assigning colors to the vertices of a graph (Figure 1 shows a graph with 7 vertices) such that no two adjacent vertices have the same color. It has many practical applications, including scheduling, resource allocation, and network design, and has been the subject of much research in both classical and quantum computing. Note that solving one NP-complete is equivalent to solving all hundreds of NP-complete problems.

The **Vertex Coloring problem** entails a set of constraints: within a graph, no two directly connected vertices should share the same color. In this context, a solution within an $n$-vertex graph is presented as a vector comprising $n$ elements, where each element corresponds to a color associated with a specific vertex in the graph. A solution is deemed *feasible* when it satisfies all constraints, ensuring that no two adjacent vertices share the same color. Conversely, a solution is classified as *infeasible* if it violates at least one of these constraints.

Several quantum approaches have been proposed to tackle the vertex coloring problem and other NP-complete problems, each offering different computational advantages. Paper [66] proposed a quantum method that has a complexity of $O(1.9140^n)$ in solving a $k$-colorable $n$-vertex graph coloring problem, demonstrating a sub-exponential improvement over classical brute-force methods but still exhibiting exponential scaling. Paper [67] presented a method based on Grover's algorithm to obtain a quadratic speedup in solving NP-complete problems.

To solve the vertex coloring problem more efficiently, this paper proposes a quantum feasibility labeling (QFL) algorithm, which introduces a novel way of classifying feasible and infeasible solutions at the quantum level. Unlike existing quantum search-based techniques that rely solely on Grover's speedup, QFL directly labels each potential solution with a feasibility indicator. This structured approach allows the VQS algorithm to efficiently identify feasible solutions, leveraging the feasibility labels as input.

A key advantage of the QFL + VQS framework over other quantum approaches is that it does not require an explicit quantum oracle to mark solutions, as the feasibility labeling is inherently embedded within the algorithm itself. This is a significant improvement over Grover-based techniques, which rely on an external oracle to define the search space, potentially increasing the overall circuit complexity. Moreover, the QFL + VQS approach scales in polynomial time (verified up to 26 qubits) with respect to the number of qubits required for feasibility labeling. If the VQS is proven to be efficient for more than 26 qubits [29], the combination of the QFL and VQS will be the first quantum algorithm to solve an NP-complete problem in polynomial time.

Additionally, compared to other existing methods:

- Classical heuristic approaches, such as DSatur (Degree of Saturation) and Backtracking, can often find approximate solutions efficiently for small to medium-sized graphs but still face exponential worst-case complexity.
- Quantum Approximate Optimization Algorithm (QAOA) has been explored for combinatorial optimization problems, but its efficiency for large-scale graph coloring remains an open question due to variational parameter optimization challenges.

- QFL + VQS offers an alternative quantum approach that leverages structured feasibility labeling and quantum search simultaneously, presenting a distinct advantage in terms of problem representation and solution retrieval.

Section II describes the QFL algorithm. Section III explains how to use VQS to find the feasible solutions to the vertex coloring problem. Results are presented in Section IV followed by conclusions given in Section V.

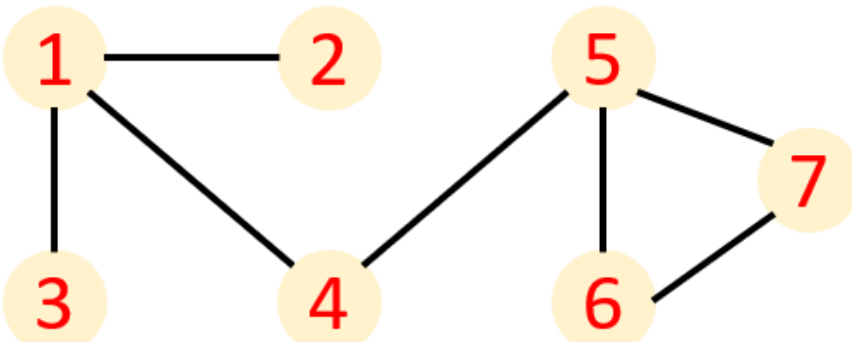

FIGURE 1. A graph with 7 vertices and 7 edges. A line between two vertices indicates the two vertices are directly connected, and we also say that the two vertices are adjacent.

## II. QUANTUM FEASIBILITY LABELING ALGORITHM

In this section, we detail the QFL algorithm. The section is structured as follows. Subsections II-A and II-B describe two pivotal components within the QFL: the SO/SOR module and the reset circuit, respectively. Subsection II-C offers a comprehensive overview of the entire quantum circuit employed in the QFL and describes the steps to construct the circuit for the QFL. Subsections II-D and II-E expand on the quantum subtraction and quantum OR circuit, illustrating their extension to accommodate additional qubits. Finally, in Subsection II-F, we conduct a detailed complexity analysis for the QFL.

The QFL generates a feasibility label for each solution. The ***feasibility label*** is represented by a single qubit which is in either $|0\rangle$ or $|1\rangle$ state, indicating that the solution is infeasible or feasible, respectively. The number of ***data qubits*** is set to the product of the number of vertices and the number of qubits per vertex.

The number of possible colors for a vertex is $k$. Then, the total number of possible solutions to the vertex coloring with $n$ vertices is equal to $k^n$. Let $m$ represent the minimal integer number such that $2^m \geq k$. Then $m \times n$ qubits represent $2^{mn}$ states which can represent all the $k^n$ solutions as $2^{mn} \geq k^n$. Therefore, a circuit of $m \times n$ qubits is used to represent all the possible solutions. The QFL uses a single qubit to represent the feasibility label of all possible solutions, which is possible because $2^{mn+1} \geq k^n + k^n$, i.e., adding one more qubit can represent the feasibility label for every possible solution.

Note that each color is uniquely represented by a unique value of a state. Take $m$=2 for example, a 2-qubit state has four possible measurement values (i.e., $|00\rangle$, $|01\rangle$, $|10\rangle$, and $|11\rangle$) and we can let them represent four different colors, respectively.

### A. QUANTUM SO/SOR MODULE

Before delving into the QFL, we detail its core module, SO/SOR. This module consists of three key operations:

1) Quantum Subtraction – Determines the difference between two quantum-encoded values.
2) Quantum OR Operation – Produces a feasibility label based on the subtraction results.
3) Reset Circuit – Resets ancilla qubits to reuse resources efficiently.

These three steps collectively form the SO/SOR module, which plays a crucial role in labeling feasible and infeasible solutions in QFL. Below, we break down each step with explanations and equations to make the process more intuitive.

The **first step** in the module is a ***quantum subtraction circuit***, which takes two groups of $m$-qubit data (denoted as $\boldsymbol{a}$ and $\boldsymbol{b}$) and computes their difference. The output of quantum subtraction is stored in $m$+1 ***Ancilla qubits***. The subtraction operation can be described as Eq. (1):

$$|\boldsymbol{a}, \boldsymbol{b}, \boldsymbol{0}, 0\rangle \rightarrow |\boldsymbol{a}, \boldsymbol{b}, \boldsymbol{a}-\boldsymbol{b}, 0\rangle \quad (1)$$

where $\boldsymbol{a}$ and $\boldsymbol{b}$ are multi-qubit number representing two values to be compared, ancilla qubits (initialized as $|\boldsymbol{0}\rangle$) store the subtraction result. In this paper, a notation in bold indicates that it involves multiple qubits.

The blue part of Figure 2 shows an example with $m$=2. As another example, if we have two 2-qubit numbers: $a=|10\rangle=2$, $b=|01\rangle=1$. Then the quantum subtraction circuit computes 2−1=1 and stores $|01\rangle$ in the ancilla qubits.

In essence, this operation checks if two values are different and prepares the result for the next step.

After performing subtraction, the **next step** is to determine whether the two values are different by analyzing the subtraction result. This is done using a ***quantum OR operation***, which checks whether at least one bit of the subtraction result is **nonzero**. If any bit is 1, it means $a \neq b$, and the feasibility label qubit should be set accordingly. A quantum OR operation can be described as Eq. (2):

$$|\boldsymbol{a}, \boldsymbol{b}, \boldsymbol{a}-\boldsymbol{b}, 0\rangle \rightarrow |\boldsymbol{a}, \boldsymbol{b}, \boldsymbol{a}-\boldsymbol{b}, d\rangle \quad (2)$$

where $d$ is a single qubit storing the logical OR of all bits in $\boldsymbol{a}-\boldsymbol{b}$, which can be described as Eq. (3):

$$d = (a-b)_1 | (a-b)_2 | \cdots | (a-b)_m | (a-b)_{m+1} \quad (3)$$

where $(a-b)_i$, $i = 1,2,\cdots,m,m+1$, represent the $i$th bit of $\boldsymbol{a}-\boldsymbol{b}$. If $d$=1, it means $a \neq b$, so the solution is **feasible**. If $d$=0, it means $a=b$, so the solution is **infeasible**.

The pink part of Figure 2 shows an example with three input qubits and one output qubit. To further illustrate subtraction in a quantum context, consider the following independent examples. When $a$=3 (represented as $|11\rangle$) and $b$=3 ($|11\rangle$), the subtraction result is 3−3=0 ($|00\rangle$), so $d$=0. If $a$=2 ($|10\rangle$) and $b$=1 ($|01\rangle$), the subtraction result is 2−1=1 ($|01\rangle$), so $d$=1.

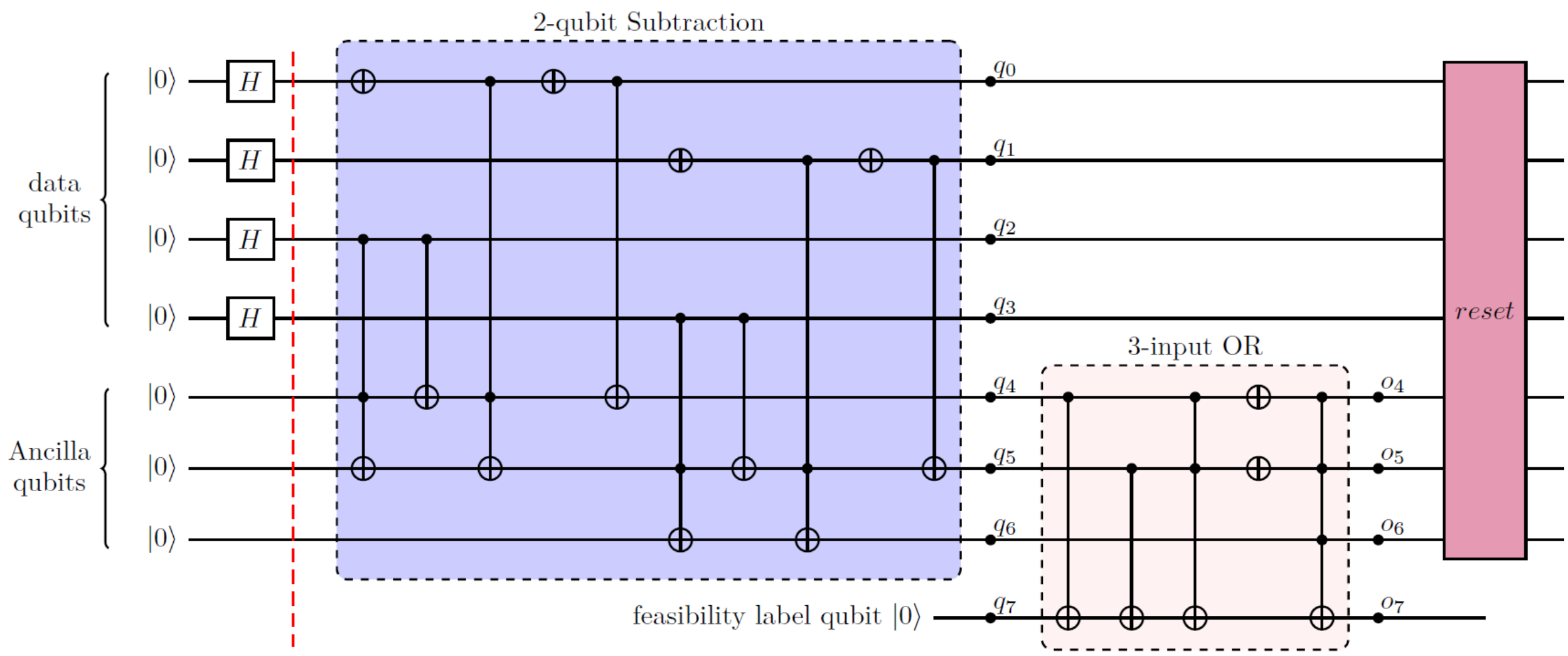

**FIGURE 2. Data input (left-hand side of the red dashed line) and an SOR module (right-hand side of the red dashed line) consisting of a 2-qubit subtraction, a 3-input OR, and a reset circuit.**

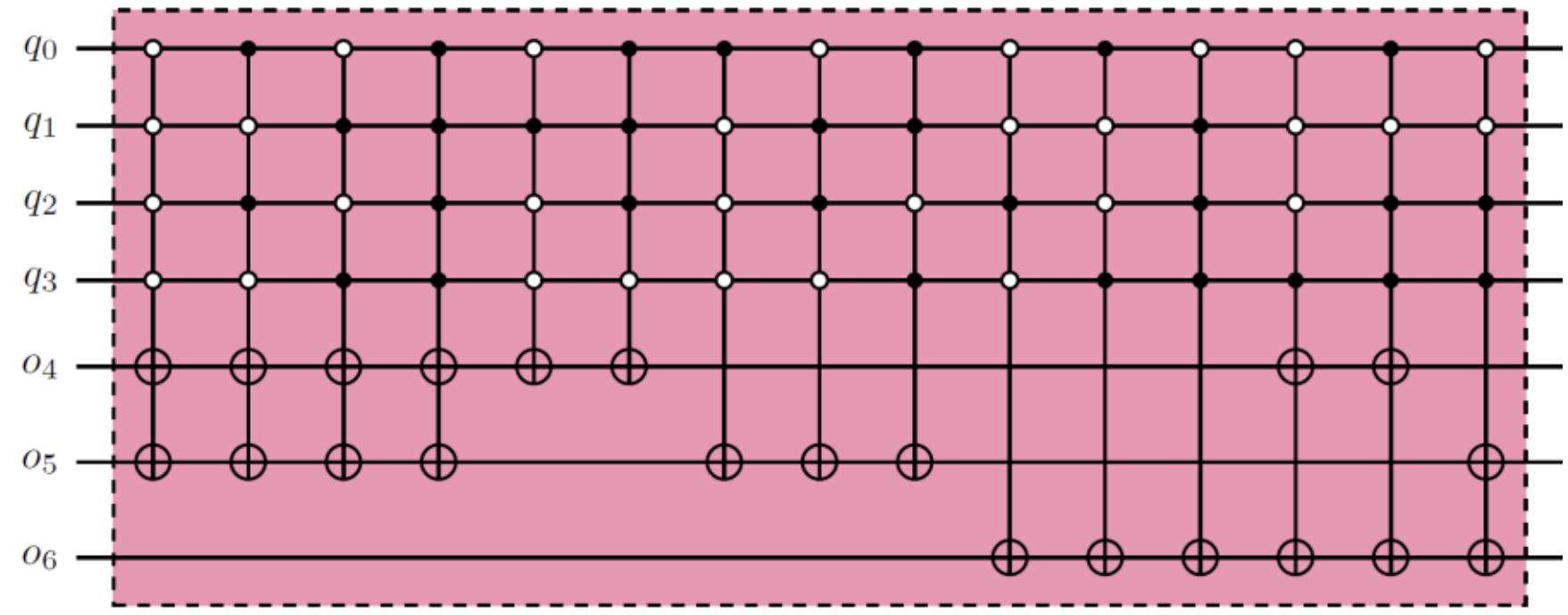

**FIGURE 3. The reset circuit used in Figure 2.**

Thus, the quantum OR gate helps determine whether a valid assignment exists based on the feasibility condition.

The **final step** in the SO/SOR module is the ***reset circuit***, which ensures that ancilla qubits used in subtraction return to their original state ($|\mathbf{0}\rangle$). This is beneficial as reusing ancilla qubits helps minimize resource usage in a quantum computation. The quantum reset can be represented as Eq. (4):

$$|\boldsymbol{a}, \boldsymbol{b}, \boldsymbol{a}-\boldsymbol{b}, d\rangle \rightarrow |\boldsymbol{a}, \boldsymbol{b}, \mathbf{0}, d\rangle \qquad (4)$$

Here, the subtraction result stored in the ancilla qubits is reset to zero, while the feasibility qubit $d$ is retained. The details of the quantum reset circuit are explained in the next subsection. An example for the 2-qubit subtraction [68] and 3-input OR module is shown in Figure 3.

For the convenience of expression, the quantum subtraction and quantum OR circuits are collectively referred to as an ***SO module***. The SO module and the reset circuit are collectively referred to as an ***SOR module***. Figure 2 shows an SOR module with data input on the far left.

To better understand the SO module, Table I provides its truth table. Note that the subtraction circuit does not change the states of input data, which can be easily verified from Figs. 2 and 5. From Table I, we can conclude that output $o_7 = 0$ when $q_3 q_2$ is equal to $q_1 q_0$ and $o_7 = 1$ when $q_3 q_2$ is different from $q_1 q_0$. This is what we need, i.e., the feasibility label is 1 when the colors of two directly connected vertices are different and 0 otherwise.

TABLE I
THE TRUTH TABLE FOR THE CIRCUIT GIVEN IN FIGURE 2, WHERE $o_7 = q_6 | q_5 | q_4$ AND $q_6 q_5 q_4 = q_1 q_0 - q_3 q_2$. NOTE: $o_6 = q_6, o_5 = \overline{q_5}, o_4 = \overline{q_4}$, WHERE THE QUANTUM STATES GIVEN IN THE TOP ROW OF THE TABLE ARE INDICATED IN FIGURE 2 (ON THE RIGHT-HAND SIDES OF THE BLUE AND PINK BLOCKS).

| row | $o_7$ | $o_6$ | $o_5$ | $o_4$ | $q_6$ | $q_5$ | $q_4$ | $q_3$ | $q_2$ | $q_1$ | $q_0$ |
|---|---|---|---|---|---|---|---|---|---|---|---|
| 1 | 0 | 0 | 1 | 1 | 0 | 0 | 0 | 0 | 0 | 0 | 0 |
| 2 | 0 | 0 | 1 | 1 | 0 | 0 | 0 | 0 | 1 | 0 | 1 |
| 3 | 0 | 0 | 1 | 1 | 0 | 0 | 0 | 1 | 0 | 1 | 0 |
| 4 | 0 | 0 | 1 | 1 | 0 | 0 | 0 | 1 | 1 | 1 | 1 |
| 5 | 1 | 0 | 0 | 0 | 0 | 1 | 1 | 0 | 0 | 1 | 1 |
| 6 | 1 | 0 | 0 | 1 | 0 | 1 | 0 | 0 | 0 | 1 | 0 |
| 7 | 1 | 0 | 0 | 1 | 0 | 1 | 0 | 0 | 1 | 1 | 1 |
| 8 | 1 | 0 | 1 | 0 | 0 | 0 | 1 | 0 | 0 | 0 | 1 |
| 9 | 1 | 0 | 1 | 0 | 0 | 0 | 1 | 0 | 1 | 1 | 0 |
| 10 | 1 | 0 | 1 | 0 | 0 | 0 | 1 | 1 | 0 | 1 | 1 |
| 11 | 1 | 1 | 0 | 0 | 1 | 1 | 1 | 0 | 1 | 0 | 0 |
| 12 | 1 | 1 | 0 | 0 | 1 | 1 | 1 | 1 | 0 | 0 | 1 |
| 13 | 1 | 1 | 0 | 0 | 1 | 1 | 1 | 1 | 1 | 1 | 0 |
| 14 | 1 | 1 | 0 | 1 | 1 | 1 | 0 | 1 | 0 | 0 | 0 |
| 15 | 1 | 1 | 0 | 1 | 1 | 1 | 0 | 1 | 1 | 0 | 1 |
| 16 | 1 | 1 | 1 | 0 | 1 | 0 | 1 | 1 | 1 | 0 | 0 |

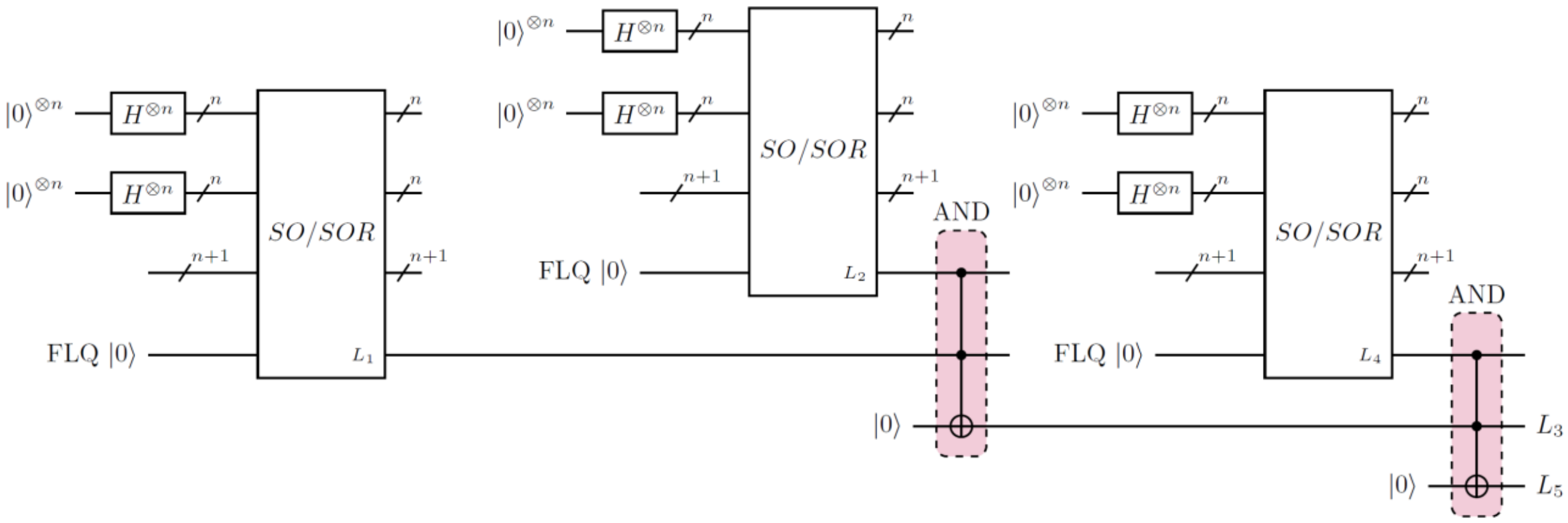

**FIGURE 4. The quantum circuit for the Quantum Feasibility Labeling algorithm, where FLQ is short for feasibility label qubit.**

### B. QUANTUM RESET CIRCUIT

Each $m$-qubit SO/SOR module needs $m$+1 Ancilla qubits, which serve as temporary storage during quantum operations. Initially, these ancilla qubits are set to $|\mathbf{0}\rangle$. However, after the module performs its computations, the ancilla qubits retain residual values that can interfere with subsequent operations.

To reuse these ancilla qubits in the next computation, we must reset them back to $|\mathbf{0}\rangle$. This is accomplished using a reset circuit. In this paper, the reset circuit uses a multi-controlled, multi-target CNOT gate to conditionally flip the state of the ancilla qubits back to $|\mathbf{0}\rangle$ based on their previous values.

The reset circuit relies on the relationship between input data qubits and the outputs of the OR circuit. This relationship is captured in Table I, which lists how the output qubits $o_4$ to $o_6$ depend on the input qubits $q_0$ to $q_3$. If an ancilla qubit ($o_4$-$o_6$) is in state $|1\rangle$, a CNOT operation is applied to flip it back. The CNOT gate is controlled by input qubits (e.g., $q_0, q_1, q_2, q_3$). Figure 3 provides an example reset circuit for the reset block (rightmost, in red) in Figure 2. Figure 3 has 15 layers, each associated with a row in Table I. Note that row 5 of Table I does not need a layer in Figure 3 because its $o_4$-$o_6$ are already in state $|\mathbf{0}\rangle$. That is, if an ancilla qubit is already in state $|0\rangle$, no reset operation is needed.

### C. ENTIRE QUANTUM CIRCUIT FOR QUANTUM FEASIBILITY LABELING

Each SO/SOR module corresponds to a constraint of the vertex coloring problem. Each SO/SOR module has a one-qubit feasibility label, denoted as $L_i$ (shown in the lower-right corner of each SO/SOR block in Figure 4), indicating whether the constraint is feasible or not. All constraints need to be satisfied for a feasible solution. Therefore, we use an AND circuit to combine two SO/SOR together in a sequential way. The feasibility label qubits of the first two SO/SOR modules are the input of the first AND circuit. Starting from the third SO/SOR module, the output of the previous AND circuit is used as the second input of the current AND circuit. According to Eqs. (1)-(4) and by denoting the data input of the $i$th SO/SOR module as $\boldsymbol{a}_i, \boldsymbol{b}_i$, we can represent the $i$th SO/SOR module as Eq. (5):

$$|\boldsymbol{a}_i, \boldsymbol{b}_i, \mathbf{0}, 0\rangle \rightarrow |\boldsymbol{a}_i, \boldsymbol{b}_i, \boldsymbol{a}_i - \boldsymbol{b}_i, d_i\rangle \quad (5)$$

where $d_i \in \{0,1\}$.

Then the input-output logic for the first AND circuit shown in Figure 4 can be expressed as Eq. (6):

$$|\boldsymbol{a}_1, \boldsymbol{b}_1, \mathbf{0}, \boldsymbol{a}_2, \boldsymbol{b}_2, \mathbf{0}, d_1, d_2, 0\rangle \rightarrow |\boldsymbol{a}_1, \boldsymbol{b}_1, \mathbf{0}, \boldsymbol{a}_2, \boldsymbol{b}_2, \mathbf{0}, d_1, d_2,\ d_1 \& d_2\rangle \quad (6)$$

where $d_1$ and $d_2$ represent the output of the feasibility label qubit of the first and second SO/SOR modules, respectively, and $d_1 \& d_2$ represents the logic AND of $d_1$ and $d_2$.

Furthermore, the input-output logic for the ($j$−1)th AND circuit, which is located at the right-hand-side of the $j$th SO/SOR module, can be represented as Eq. (7):

$$|\boldsymbol{A}, \boldsymbol{B}, \mathbf{0}, d_j, D_{j-1}, 0\rangle \rightarrow |\boldsymbol{A}, \boldsymbol{B}, \mathbf{0}, d_j, D_{j-1},\ D_j\rangle,\ 3 \le j \le g \quad (7)$$

where $\boldsymbol{A}, \boldsymbol{B}, \mathbf{0}$ represent all the data input and state $|0\rangle$ that is input into all ancilla qubits, $g$ denotes the number of edges in the coloring problem, $d_j$ represents the output of the feasibility label qubit of the $j$th SO/SOR modules, respectively, $D_{j-1}$ represents $d_1 \& d_2 \& \cdots \& d_{j-1},\ j \ge 3,$ and $D_j$ represents $D_{j-1} \& d_j$ which can be expanded into $D_j = d_1 \& d_2 \& \cdots \& d_{j-1} \& d_j$. For a $g$-edge coloring problem, there are $g$ SO/SOR module and, therefore, the maximum value of $j$ is $g$.

For the data input, each vertex has $m$ qubits. If a vertex is involved in multiple SO/SOR blocks, all blocks should use the same $m$ qubits for that vertex, which is possible because the states of the data qubits do not change before and after the SO/SOR block. Also, the reset circuit does not change the state of any data qubits.

We describe the construction of the entire circuit of QFL in the following 4 steps.

Step 1 - **Feasibility label**: Use two SO/SOR modules (see the left and middle parts of Figure 4) to obtain two feasibility labels (i.e., $L_1$ and $L_2$ in Figure 4) for the first two constraints, respectively.

Step 2 - **Quantum AND**: The two feasibility labels are converted into one feasibility label using a quantum AND circuit (see the red part in the middle of Figure 4). The output of this circuit is a one-qubit feasibility label ($L_3$ in Figure 4). The logical relationship for this AND circuit can be represented as $L_3$=$L_1$&$L_2$.

Step 3 - **Incremental step**: Then, a new SOR circuit (see the right side of Figure 4) is added for a new constraint of the vertex coloring problem. The output at the feasibility label qubit of this circuit ($L_4$) and the feasibility label ($L_3$) obtained from the previous AND circuit are used as inputs to a new quantum AND circuit (see the rightmost red part of Figure 4). The output of the new quantum AND circuit is a one-qubit feasibility label ($L_5$). The logical relationship for this circuit can be expressed as $L_5$=$L_3$&$L_4$.

Step 4 - **Repeat the incremental step** for each remaining constraint of the vertex coloring problem.

Now, the feasibility label qubit from the last quantum AND circuit and the data qubits together represent all solutions and their feasibility labels. For an $n$-vertex $g$-edge coloring problem, there are $g$ modules, and the $D_j$ in Eq. (7), where $j = g$, is the feasibility label representing the feasibility of all solutions $\boldsymbol{a}_1, \boldsymbol{b}_1, \boldsymbol{a}_2, \boldsymbol{b}_2, \cdots, \boldsymbol{a}_n, \boldsymbol{b}_n$.

### D. QUANTUM SUBTRACTION WITH MORE QUBITS

This subsection presents the $m$-qubit quantum subtraction circuit, shown in Figure 2 (blue block) and Figure 5 for $m$=2 and $m$=3, respectively. In Figure 5, the logical relationship of the 3-qubit quantum subtraction can be expressed as $q_9q_8q_7q_6 = q_2q_1q_0 - q_5q_4q_3$. Then, an $m$-qubit quantum subtraction circuit can be derived, i.e., connecting $m$ SUBT blocks in series such that the indices of the four qubits connected to the four wires of a block differ from those of its adjacent block by exactly 1, respectively, where $m \geq 2$.

### E. QUANTUM OR CIRCUITS WITH MORE QUBITS

This subsection presents the $m$-qubit quantum OR circuit, shown in Figure 2 (pink block) and Figure 6 for $m$=3, and $m$=4, respectively. In Figure 6, the logical relationship of the 4-qubit quantum OR can be expressed as $q_4 = q_0|q_1|q_2|q_3$. Then, an $m$-input quantum OR circuit can be derived, and it can be constructed using the pseudo code given in Algorithm I.

### F. COMPLEXITY ANALYSIS

This subsection calculates the number of qubits and the depth of the circuit required by the QFL for solving a $k$-colorable $n$-vertex $g$-edge coloring problem.

In the calculation given in the following two paragraphs, the reset circuit (the red block on the far right in Figure 2) is used, i.e., each SO/SOR module in Figure 4 uses SOR. The total depth of the entire circuit is $g\left[5m+1+\left(\sum_{i=1}^{m}\binom{m}{i}+2\right)+(2^{2m}-1)\right]$, where the four terms in the square brackets are the depths of the subtraction, AND, OR, and reset circuits, respectively, and $\binom{m}{i}$ denotes the number of $i$-combinations from $m$ elements. Note that $\sum_{i=1}^{m}\binom{m}{i} = 2^m - 1$. Then, the

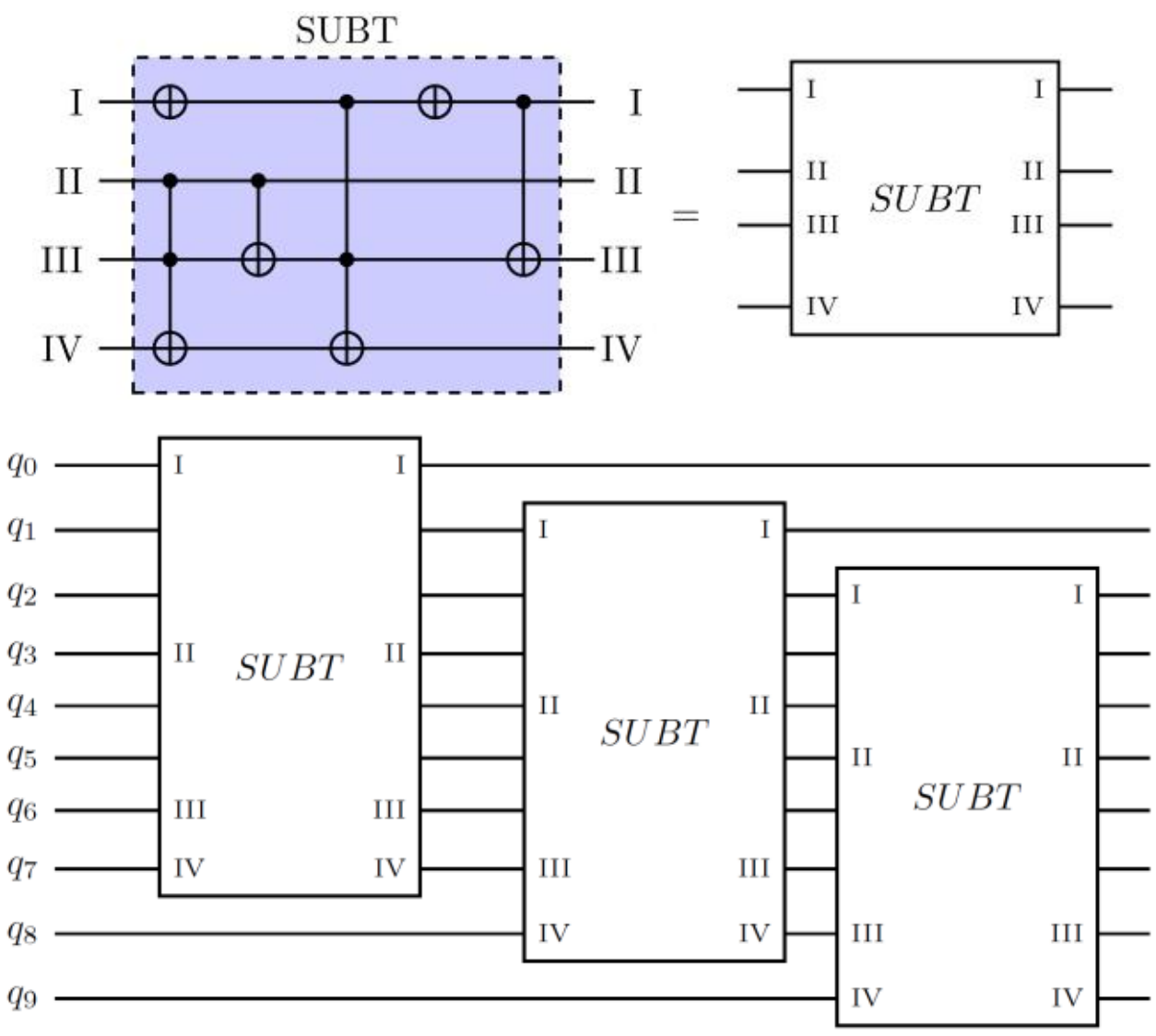

**FIGURE 5.** Quantum subtraction circuits. The top left panel in blue is a 1-qubit quantum subtraction circuit having four qubits indicated by I, II, III, and IV, respectively. Its abstract form (denoted as SUBT) is shown in the top right panel. The bottom panel shows a 3-qubit quantum subtraction circuit consisting of three 1-qubit quantum subtraction circuits. The I, II, III, and IV indicate where the four qubits of a SUBT block are connected to $q_0$-$q_9$. For example, the four qubits of the leftmost SUBT block are connected to $q_0$, $q_3$, $q_6$, and $q_7$, respectively.

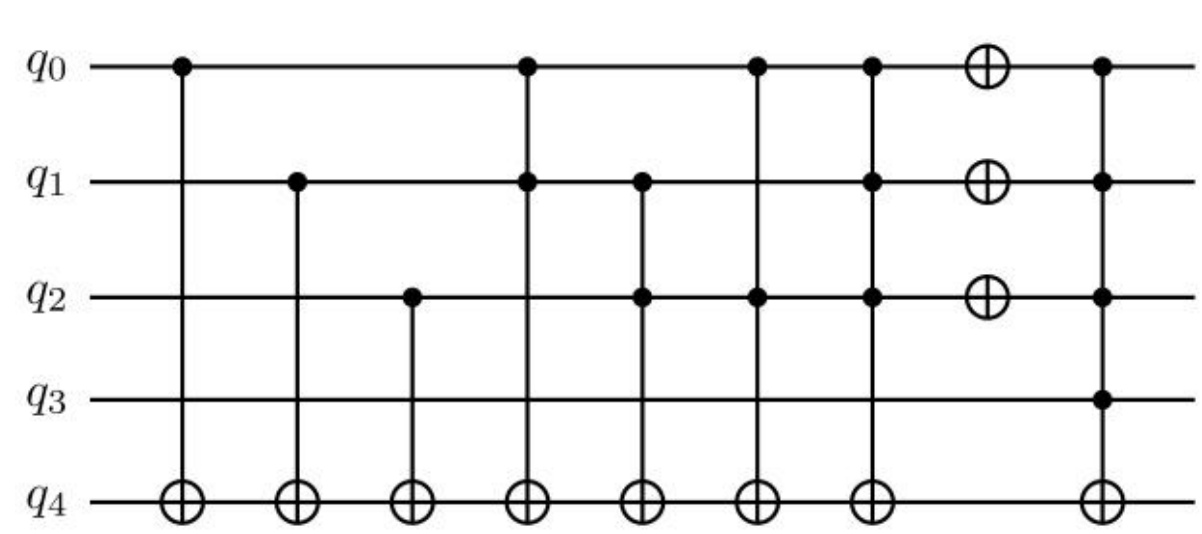

**FIGURE 6.** A 4-input quantum OR circuit.

ALGORITHM I
PSEUDO CODE FOR GENERATING THE $m$-INPUT QUANTUM OR CIRCUIT.

**Input**: $m$+1 qubits
**Output**: $m$-input quantum OR circuit

| | |
|---|---|
| 1 | Add $m$−1 CNOT gates with the control qubits located at qubits $q_0$~$q_{m-2}$, respectively, and the target qubit for each gate is located at qubit $q_{m+1}$. |
| 2 | Let $u$=2 |
| 3 | **while** $u \leq m-1$ |
| 4 | Add a $C^u(X)$ gate with $u$ control qubits located at set $\mathcal{L}$ and a target qubit located at qubit $q_{m+1}$. |
| 5 | Repeat the previous step for $\mathcal{L}$ being any combinations of $u$ qubits from $q_0$ to $q_{m-2}$. |
| 6 | $u \leftarrow u$+1. |
| 7 | Add a Pauli X gate to each of qubits $q_0$~$q_{m-2}$. |
| 8 | Add a $C^m(X)$ gate with a target qubit located at qubit $q_{m+1}$ and $m$ control qubits located at qubits $q_0$~$q_{m-1}$. |

total depth can be simplified to $g(2^{2m}+2^m+5m+1)$. Note that $m$ is related to the number of possible colors $k$, i.e., $m=\lceil\log_2 k\rceil$, where the symbol $\lceil\cdot\rceil$ means rounding up to the nearest integer number. Then the total depth can be written as $O(k^2 g)$.

The total number of qubits is $mn+(m+1)+g+(g-1)$ where the four terms correspond to the numbers of data qubits, Ancilla qubits, feasibility label qubits, output qubits of quantum AND circuits, respectively. It can be simplified as $mn+m+2g$.

In the calculation given in this paragraph, the reset circuit is not used, i.e., each SO/SOR module in Figure 4 uses SO. The total depth of the entire circuit is $g\left[5m+1+\left(\sum_{i=1}^{m}\binom{m}{i}+2\right)\right]$ which can be simplified to $g[5m+2^m+2]$. Then the total depth can be written as $O(kg)$. As the reset circuit is not used, each SO module requires new Ancilla qubits. Therefore, the total number of qubits is $mn+(m+1)g+g+(g-1)$ which can be rewritten as $mn+(m+3)g-1$.

Table II lists the circuit depths and the numbers of qubits in both situations, showing that using the reset circuit reduces the required number of qubits at the expense of deeper circuit depth. When $k$ (the number of colors) is large, we recommend not using the reset circuit. On the other hand, when $g$ (the number of edges) is large and $k$ is small, it is better to use the reset circuit.

TABLE II
THE CIRCUIT DEPTHS AND THE NUMBERS OF QUBITS REQUIRED BY THE QFL WITH AND WITHOUT USING THE RESET CIRCUIT FOR SOLVING A $k$-COLORABLE $n$-VERTEX $g$-EDGE COLORING PROBLEM, WHERE $m=\lceil\log_2 k\rceil$.

| With Reset Circuit | Circuit Depth | # of qubits |
| --- | --- | --- |
| Yes | $O(k^2 g)$ | $mn+m+2g$ |
| No | $O(kg)$ | $mn+(m+3)g-1$ |

## III. VARIATIONAL QUANTUM SEARCH

In this section, we discuss how to utilize the VQS to find all feasible solutions to the vertex coloring problem using the output of the QFL as the input to the VQS.

Within the QFL output, there exists a crucial component known as the final labeling qubit, denoted as $D_j$ in Eq. (7). This qubit assumes a state of $|1\rangle$ to represent a feasible solution and $|0\rangle$ to denote an infeasible one. When combined with all possible solutions, this labeling qubit collectively forms an unstructured database. Each element in this database corresponds to a solution, accompanied by its feasibility label. A feasible solution is analogous to a good element, whereas an infeasible solution corresponds to a bad element. This conceptual framework effectively links feasible and infeasible solutions to the concept of an unstructured database. In other words, the QFL algorithm converts the vertex coloring problem into an unstructured database where good elements are labeled with qubit $|1\rangle$ and bad elements are labeled with qubit $|0\rangle$.

Grover's search algorithm (GSA) applies a negative phase to a good element such that the GSA can increase the probability of finding it. In contrast, VQS takes a different approach by attaching a label state, $|1\rangle$, to the good elements and $|0\rangle$ to the bad elements. Both GSA and VQS employ this strategy to amplify the likelihood of identifying the good elements within an unstructured database, as described in [29].

Considering the feasible and infeasible solutions are analogous to good and bad elements within an unstructured database, as described above, VQS can amplify the probability of identifying the feasible solutions. In other words, VQS leverages the label state $|1\rangle$ on the final labeling qubit to amplify the probability of finding feasible solutions, much like it amplifies the probability of identifying good elements in an unstructured database. Since the QFL algorithm already provides the label qubit, the oracle component in the VQS is not needed when searching for the unstructured database generated by the QFL.

## IV. SCALABILITY CHALLENGES OF THE QFL + VQS FRAMEWORK

Extending the QFL + VQS framework to larger quantum systems introduces several scalability challenges. Below, we outline key challenges and discuss potential solutions to improve scalability.

One of the primary challenges in scaling QFL + VQS is the limited connectivity between qubits on current quantum hardware. Most near-term quantum processors have restricted qubit connectivity, meaning that two-qubit operations (e.g., CNOT gates) can only be applied between specific pairs of qubits. As the number of qubits increases, the need for swap gates to route information between distant qubits grows, leading to increased circuit depth and noise accumulation. One potential solution is to employ advanced qubit mapping techniques to reduce the number of swap operations and improve gate efficiency.

Another challenge arises from the optimization of parameterized quantum circuits within the VQS algorithm. As the problem size increases, VQS optimization becomes more difficult due to barren plateaus, where gradients vanish and hinder effective training. Potential solutions include optimizing VQS parameters layer by layer instead of training the entire circuit at once and using advanced methods to provide a good initialization of variational parameters, which could reduce the optimization search space and improve convergence.

Furthermore, as the number of vertices, edges, and colors in the vertex coloring problem increases, both circuit depth and qubit count grow correspondingly. Larger circuits introduce more gate errors due to noise and decoherence, limiting the accuracy of results. To mitigate this, error suppression and mitigation techniques can be employed to reduce the impact of gate errors. Additionally, as fault-tolerant quantum computing continues to advance, quantum error correction methods will play a crucial role in preserving computational fidelity for large-scale implementations.

## V. RESULT

### A. Results of Vertex Coloring Problem

We have implemented the QFL on an IBM Qiskit [34] simulator to solve a 4-colorable 4-vertex 3-edge coloring problem. The three edges are 1-2, 1-3, and 1-4, i.e., vertex 1 connects to vertices 2, 3, and 4, respectively. Based on the principles of permutation and combination, vertex 1 can be colored with any of the four available colors, and then vertices 2, 3, and 4 can each be colored with any of the remaining three colors. As a result, the number of possible feasible solutions is $4 \times 3 \times 3 \times 3 = 108$. The circuit used by the QFL is given in Figure 4, where the three SOR modules are for the three constraints, respectively. These modules use the same qubits for vertex 1, i.e., the output of one SOR block at these qubits is used as input for the next SOR.

The circuit is run and measured 20,000 times with the results shown in Table III. The table shows that 256 quantum states are obtained, which exactly corresponds to all possible combinations of 8 qubits, i.e., $2^8$=256 (counted from the far right of each state). The feasibility label qubit is the 9th qubit counted from the right. The labels of the states shown in the first 37 rows are 0, while those in rows 38-64 are 1. A label of 0 represents an infeasible solution, while a label of 1 indicates a feasible solution. Considering that each row has 4 states, there are 148 infeasible states and 108 feasible states, which matches the number calculated in the previous paragraph and validates the QFL.

### B. Real-world Applications of the Vertex Coloring Problem

To demonstrate the practical utility of our QFL approach, combined with the VQS algorithm, in solving real-world problems, we present two examples that can be modeled as vertex coloring problems. In each case, the underlying problem is converted into a vertex coloring formulation, after which QFL + VQS is used to find a feasible coloring.

*Example 1: University Course Scheduling* -- In a university setting, scheduling courses presents a challenge when certain courses cannot be held simultaneously due to shared students or instructors. This problem can be effectively modeled as a vertex coloring problem:

1) Each course is represented as a vertex in the graph.

2) An edge is drawn between two vertices if the corresponding courses share students or instructors, indicating that they cannot be scheduled at the same time.

3) The goal is to find the minimum number of timeslots (colors) required to schedule all courses such that no two conflicting courses (connected vertices) are assigned the same timeslot (color).

Using QFL + VQS, we can efficiently determine if a given number of colors (timeslots) leads to a feasible schedule, ensuring no conflicting courses share the same timeslot.

*Example 2: Radio Frequency Assignment for Transmitters* -- In telecommunications, radio transmitters (e.g., cell towers) must operate on different frequency channels if they are in close proximity to avoid interference. This scenario can also be modeled as a vertex coloring problem:

1) Each transmitter is represented as a vertex in the graph.

2) An edge connects two vertices if the corresponding transmitters are close enough to interfere with each other.

3) The task is to assign frequencies (colors) to transmitters (vertices) such that no two adjacent transmitters share the same frequency, while minimizing the total number of frequencies used.

Here, QFL + VQS can efficiently verify if a given set of frequencies could accommodate all transmitters without interference, ensuring no adjacent transmitters share the same frequency.

### C. Applying QFL+VQS to Solve the SAT Problem

In this section, we extend the application of the QFL + VQS framework to other NP-complete problems. One such problem is the SAT, which plays a fundamental role in areas such as artificial intelligence, hardware verification, and cryptography.

The SAT problem is defined as follows. Given a Boolean formula in Conjunctive Normal Form (CNF) (a set of clauses connected by AND operations), determine whether there exists an assignment of Boolean variables that satisfies all clauses. Each clause is a disjunction (OR operation) of literals (Boolean variables or their negations). For example, here is a three-variable, three-clause SAT problem: $(x_1 \vee \neg x_2) \wedge (\neg x_1 \vee x_3) \wedge (x_2 \vee x_3)$, where $x_1, x_2, x_3$ are Boolean variables, $\vee$ represents the logical OR operation, $\neg x_2$ denotes the negation of $x_2$. The formula is satisfied if at least one literal in each clause is true.

Now we describe how we can use QFL + VQS to solve the problem. We use a qubit to represent each Boolean variable $x_i$, where the initial state of the qubit is an equal superposition of $|0\rangle$ (representing False) and $|1\rangle$ (representing True), given by $(|0\rangle + |1\rangle)/\sqrt{2}$. We can use ancilla qubits to store clause evaluations. Then, we use a quantum circuit to apply OR gates (see Figure 2) between literals in each clause followed by using AND operations (refer to Figure 4) across all clauses to compute the final feasibility label. Subsequently, we feed the feasibility-labeled quantum state into VQS. Assume label $|1\rangle$ represent feasible, while $|0\rangle$ represents infeasible. At last, the VQS searches for all feasible solutions by amplifying their probabilities. If the SAT formula is satisfiable, VQS will find one or more valid solutions with high probability. If the formula is unsatisfiable, the feasibility qubit always collapses to $|0\rangle$, indicating no valid solution exists.

## VI. CONCLUSION

In summary, we have successfully developed a QFL algorithm that provides a feasibility label for every possible solution to the vertex coloring problem. The QFL algorithm converts the vertex coloring problem into the problem of searching an unstructured database where good and bad elements are labeled with qubits $|1\rangle$ and $|0\rangle$, respectively. The output of this algorithm can be directly input into the VQS algorithm, which can then find all feasible solutions. Notably, the computational complexity of the QFL is a polynomial function of the color options, vertices, and edges within a given vertex coloring problem.

We had numerically validated the efficiency of the VQS for cases involving up to 26 qubits in our previous work. The next critical milestone lies in extending this efficiency to all possible qubit numbers, which, if achieved, would mark a profound breakthrough. This achievement would signify that NP-complete problems could be efficiently addressed through the power of quantum algorithms.

TABLE III

Measurement results. The strings in single quotes represent measured quantum states, and the number after a colon is the number of times that the state shown in the same cell is obtained from measurement.

| row | State:count | State:count | State:count | State:count |
|---|---|---|---|---|
| 1 | '0000000000000000': 86 | '0000000000000100': 80 | '0000000000000101': 64 | '0000000000001000': 83 |
| 2 | '0000000000001010': 83 | '0000000000001100': 84 | '0000000000001111': 72 | '0000000000010000': 69 |
| 3 | '0000000000010001': 69 | '0000000000010100': 68 | '0000000000010101': 93 | '0000000000011000': 71 |
| 4 | '0000000000011001': 77 | '0000000000011010': 69 | '0000000000011100': 80 | '0000000000011101': 65 |
| 5 | '0000000000011111': 78 | '0000000000100000': 86 | '0000000000100010': 81 | '0000000000100100': 75 |
| 6 | '0000000000100101': 68 | '0000000000100110': 90 | '0000000000101000': 68 | '0000000000101010': 75 |
| 7 | '0000000000101100': 84 | '0000000000101110': 60 | '0000000000101111': 68 | '0000000000110000': 69 |
| 8 | '0000000000110011': 89 | '0000000000110100': 80 | '0000000000110101': 77 | '0000000000110111': 71 |
| 9 | '0000000000111000': 84 | '0000000000111010': 72 | '0000000000111011': 77 | '0000000000111100': 73 |
| 10 | '0000000000111111': 81 | '0000000001000000': 68 | '0000000001000001': 80 | '0000000001000100': 89 |
| 11 | '0000000001000101': 90 | '0000000001001000': 78 | '0000000001001001': 81 | '0000000001001010': 69 |
| 12 | '0000000001001100': 85 | '0000000001001101': 84 | '0000000001001111': 75 | '0000000001010000': 72 |
| 13 | '0000000001010001': 72 | '0000000001010101': 83 | '0000000001011001': 70 | '0000000001011010': 79 |
| 14 | '0000000001011101': 76 | '0000000001011111': 89 | '0000000001100000': 88 | '0000000001100001': 74 |
| 15 | '0000000001100010': 67 | '0000000001100101': 68 | '0000000001100110': 74 | '0000000001101001': 82 |
| 16 | '0000000001101010': 65 | '0000000001101101': 61 | '0000000001101110': 84 | '0000000001101111': 84 |
| 17 | '0000000001110000': 87 | '0000000001110001': 74 | '0000000001110011': 72 | '0000000001110101': 77 |
| 18 | '0000000001110111': 81 | '0000000001111001': 80 | '0000000001111010': 76 | '0000000001111011': 79 |
| 19 | '0000000001111101': 94 | '0000000001111111': 88 | '0000000010000000': 86 | '0000000010000010': 78 |
| 20 | '0000000010000100': 79 | '0000000010000101': 82 | '0000000010000110': 88 | '0000000010001000': 69 |
| 21 | '0000000010001010': 67 | '0000000010001100': 74 | '0000000010001110': 78 | '0000000010001111': 84 |
| 22 | '0000000010010000': 87 | '0000000010010001': 89 | '0000000010010010': 89 | '0000000010010101': 74 |
| 23 | '0000000010010110': 63 | '0000000010011001': 66 | '0000000010011010': 75 | '0000000010011101': 70 |
| 24 | '0000000010011110': 80 | '0000000010011111': 78 | '0000000010100000': 66 | '0000000010100010': 80 |
| 25 | '0000000010100101': 89 | '0000000010100110': 73 | '0000000010101010': 81 | '0000000010101110': 80 |
| 26 | '0000000010101111': 73 | '0000000010110000': 84 | '0000000010110010': 69 | '0000000010110011': 86 |
| 27 | '0000000010110101': 83 | '0000000010110110': 92 | '0000000010110111': 73 | '0000000010111010': 88 |
| 28 | '0000000010111011': 73 | '0000000010111110': 76 | '0000000010111111': 83 | '0000000011000000': 71 |
| 29 | '0000000011000011': 86 | '0000000011000100': 79 | '0000000011000101': 75 | '0000000011000111': 67 |
| 30 | '0000000011001000': 80 | '0000000011001010': 70 | '0000000011001011': 89 | '0000000011001100': 68 |
| 31 | '0000000011001111': 71 | '0000000011010000': 68 | '0000000011010001': 84 | '0000000011010011': 70 |
| 32 | '0000000011010101': 80 | '0000000011010111': 70 | '0000000011011001': 93 | '0000000011011010': 71 |
| 33 | '0000000011011011': 84 | '0000000011011101': 81 | '0000000011011111': 67 | '0000000011100000': 97 |
| 34 | '0000000011100010': 83 | '0000000011100011': 87 | '0000000011100101': 73 | '0000000011100110': 80 |
| 35 | '0000000011100111': 86 | '0000000011101010': 78 | '0000000011101011': 70 | '0000000011101110': 68 |
| 36 | '0000000011101111': 63 | '0000000011110000': 81 | '0000000011110011': 68 | '0000000011110101': 79 |
| 37 | '0000000011110111': 72 | '0000000011111010': 89 | '0000000011111011': 87 | '0000000011111111': 83 |
| 38 | '0000000100000001': 67 | '0000000100000010': 104 | '0000000100000011': 93 | '0000000100000110': 66 |
| 39 | '0000000100000111': 83 | '0000000100001001': 66 | '0000000100001011': 58 | '0000000100001101': 66 |
| 40 | '0000000100001110': 72 | '0000000100010010': 77 | '0000000100010011': 89 | '0000000100010110': 91 |
| 41 | '0000000100010111': 73 | '0000000100011011': 66 | '0000000100011110': 90 | '0000000100100001': 92 |
| 42 | '0000000100100011': 80 | '0000000100100111': 78 | '0000000100101001': 80 | '0000000100101011': 82 |
| 43 | '0000000100101101': 77 | '0000000100110001': 70 | '0000000100110010': 81 | '0000000100110110': 73 |
| 44 | '0000000100111001': 105 | '0000000100111101': 84 | '0000000100111110': 82 | '0000000101000010': 69 |
| 45 | '0000000101000011': 94 | '0000000101000110': 74 | '0000000101000111': 76 | '0000000101001011': 84 |
| 46 | '0000000101001110': 83 | '0000000101010010': 83 | '0000000101010011': 78 | '0000000101010100': 65 |
| 47 | '0000000101010110': 75 | '0000000101010111': 86 | '0000000101011000': 81 | '0000000101011011': 64 |
| 48 | '0000000101011100': 91 | '0000000101011110': 79 | '0000000101100011': 83 | '0000000101100100': 77 |
| 49 | '0000000101100111': 66 | '0000000101101000': 82 | '0000000101101011': 64 | '0000000101101100': 79 |
| 50 | '0000000101110010': 83 | '0000000101110100': 77 | '0000000101110110': 92 | '0000000101111000': 83 |
| 51 | '0000000101111100': 98 | '0000000101111110': 86 | '0000000110000001': 88 | '0000000110000011': 88 |
| 52 | '0000000110000111': 76 | '0000000110001001': 78 | '0000000110001011': 73 | '0000000110001101': 87 |
| 53 | '0000000110010011': 65 | '0000000110010100': 80 | '0000000110010111': 95 | '0000000110011000': 84 |
| 54 | '0000000110011011': 90 | '0000000110011100': 68 | '0000000110100001': 67 | '0000000110100011': 71 |
| 55 | '0000000110100100': 77 | '0000000110100111': 78 | '0000000110101000': 82 | '0000000110101001': 87 |
| 56 | '0000000110101011': 84 | '0000000110101100': 71 | '0000000110101101': 67 | '0000000110110001': 86 |
| 57 | '0000000110110100': 78 | '0000000110111000': 76 | '0000000110111001': 92 | '0000000110111100': 82 |
| 58 | '0000000110111101': 67 | '0000000111000001': 70 | '0000000111000010': 81 | '0000000111000110': 100 |
| 59 | '0000000111001001': 73 | '0000000111001101': 69 | '0000000111001110': 80 | '0000000111010010': 75 |
| 60 | '0000000111010100': 82 | '0000000111010110': 72 | '0000000111011000': 75 | '0000000111011100': 76 |
| 61 | '0000000111011110': 91 | '0000000111100001': 72 | '0000000111100100': 69 | '0000000111101000': 74 |
| 62 | '0000000111101001': 83 | '0000000111101100': 69 | '0000000111101101': 74 | '0000000111110001': 78 |
| 63 | '0000000111110010': 95 | '0000000111110100': 80 | '0000000111110110': 85 | '0000000111111000': 66 |
| 64 | '0000000111111001': 75 | '0000000111111100': 59 | '0000000111111101': 78 | '0000000111111110': 78 |

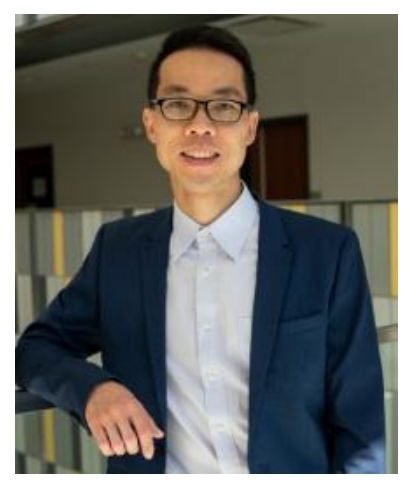

**JUNPENG ZHAN** (M'16) received the B.Eng. and Ph.D. degrees in electrical engineering from Zhejiang University, China, in 2009 and 2014, respectively.

He is currently an Assistant Professor at Alfred University, Alfred, New York, USA. He was a Research Associate Electrical Engineer with the Sustainable Energy Technologies Department, Brookhaven National Laboratory, USA. He was a Postdoctoral Fellow with the Department of Electrical and Computer Engineering, University of Saskatchewan, Saskatoon, SK, Canada, and a Visiting Research Assistant with the Department of Electrical Engineering and Electronics, University of Liverpool, U.K.

Dr. Zhan's research interests include quantum computing, smart grid modeling, planning and operation, machine learning, renewable energy, and power system optimization. He is an Associate Editor of IET Generation, Transmission & Distribution.